\documentclass[prd,10pt,aps,twocolumn,superscriptaddress]{revtex4-2}
\usepackage{graphicx}
\usepackage[T1]{fontenc}
\usepackage{lmodern}
\usepackage[dvipsnames,x11names]{xcolor}
\usepackage[colorlinks=true,linkcolor=Magenta!80!black,citecolor=Magenta!70!black,urlcolor=Magenta!80!black]{hyperref}
\usepackage{amsmath}
\usepackage{amssymb}
\usepackage{mathtools}
\usepackage{multirow}
\usepackage{orcidlink}
\usepackage{booktabs}
\usepackage{bm}
\usepackage{slashed}

\newcommand{\Dperp}{\bm{\Delta}_\perp}
\newcommand{\barm}{\bar{m}}
\newcommand{\bu}{\bar{u}}
\hypersetup{pdfencoding=auto,psdextra}

\begin{document}

\title{Chiral-odd structure of the $N \to \Delta$ transition: transverse multipoles from QCD light-cone sum rules}

\author{Ula\c{s} \"{O}zdem\orcidlink{0000-0002-1907-2894}}
\email[]{ulasozdem@aydin.edu.tr}
\affiliation{Health Services Vocational School of Higher Education, Istanbul Aydin University, Sefakoy-Kucukcekmece, 34295 Istanbul, T\"{u}rkiye}

\begin{abstract}
The matrix element of the tensor current between the nucleon and the $\Delta(1232)$ is described by six form factors, which organize into transverse multipoles of rank zero to three. The rank-three term, an octupole, has no counterpart in the diagonal tensor matrix elements of the nucleon or in the electromagnetic $N \to \Delta$ transition. We study the isovector chiral-odd transition within QCD light-cone sum rules, constructing the sum rules from light-cone projections of the correlation function rather than from individual Lorentz structures and removing the contributions of spin-$1/2$ states of both parities explicitly. At leading and next-to-leading power in the light-cone expansion the spin-$1/2$-free sum rules form an overdetermined system that determines four combinations of the form factors, among them the transverse dipole, quadrupole and octupole; the monopole and the longitudinal form factor $F_1$ are not accessible at this level. Conventional sum rules built on single Lorentz structures that measure the same combination of form factors differ by up to a factor of two while being equally stable in the Borel parameter, and we show that these structures receive no contribution of leading or next-to-leading power. With the $\Delta$ coupling and the continuum threshold taken from the $\Delta$ two-point sum rule and two sets of nucleon distribution amplitudes, we obtain the quadrupole for $1 \leq Q^2 \leq 6$~GeV$^2$ and the octupole for $Q^2 \gtrsim 2$~GeV$^2$, with Borel and threshold variations of at most about ten percent, and the dipole, whose magnitude depends strongly on the distribution amplitudes. The model of the distribution amplitudes is the dominant source of uncertainty, and the results do not determine an extrapolation to $Q^2 = 0$.
\end{abstract}

\maketitle

\section{Introduction}
\label{sec:introduction}

Among the local operators whose matrix elements probe hadronic structure, the tensor current $\bar q\,\sigma_{\mu\nu}\,q$ occupies a distinctive position. Unlike the vector and axial-vector currents that underlie the electromagnetic and weak form factors, the tensor current is chiral-odd: it connects quark fields of opposite chirality, and its matrix elements therefore encode information about the transverse spin of quarks inside hadrons~\cite{Ralston:1979ys, Artru:1989zv, Jaffe:1991kp, Jaffe:1991ra, Cortes:1991ja, Barone:2001sp}. This information is not accessible through inclusive deep inelastic scattering, and chiral-odd observables --- transversity distributions, tensor charges, and the corresponding generalized parton distributions (GPDs) --- have to be reconstructed from semi-inclusive and hard exclusive processes~\cite{Anselmino:2007fs, Anselmino:2008jk, Anselmino:2013vqa}. The diagonal nucleon tensor charges have been computed in bag models~\cite{He:1994gz}, light-front and constituent quark models~\cite{Pasquini:2005dk, Gamberg:2001qc}, the external-field approach~\cite{He:1996wy}, and lattice QCD~\cite{LHPC:2007blg, QCDSF:2006tkx, Bhattacharya:2015wna}, and the tensor form factors are tied to the chiral-odd GPDs through their first Mellin moments~\cite{Diehl:2001pm, Diehl:2003ny, Belitsky:2005qn}.

The $\Delta(1232)$ resonance is a natural setting in which to extend this program. The electromagnetic $N \to \Delta$ transition has been measured precisely at JLab, MAMI and ELSA~\cite{Aznauryan:2011qj, Pascalutsa:2006up, CLAS:2009ces}, and these data have constrained quark-model, lattice and chiral-effective-theory descriptions~\cite{Alexandrou:2008bn, Alexandrou:2010jv, Boinepalli:2009sq, Geng:2009ys, Li:2016ezv, Krivoruchenko:1991pm, Berger:2004yi, Schlumpf:1993rm, Pascalutsa:2007wz, Aliev:2009pd, Lee:1997jk, Azizi:2009egn}. The axial-vector sector has been mapped within QCD light-cone sum rules (LCSR)~\cite{Aliev:2007pi, Kucukarslan:2015urd} and lattice QCD~\cite{Alexandrou:2006mc}, and the gravitational transition has been parametrized~\cite{Kim:2022bia} and computed with LCSR~\cite{Ozdem:2022zig}. The chiral-odd sector of the same transition has, to our knowledge, not been addressed quantitatively, although the diagonal tensor sector is well developed for the nucleon~\cite{Erkol:2011iw, Aliev:2011ku, Gutsche:2016xff, Azizi:2019ytx, Ozdem:2020vpt, Ozdem:2021zbn}, the octet~\cite{kucukarslan:2016xhx} and the decuplet~\cite{Asmaee:2025elo, Asmaee:2026wrk}. The classification of the $N \to \Delta$ transition GPDs, revisited from the large-$N_c$~\cite{Goeke:2001tz} and partial-wave~\cite{Kim:2024hhd} perspectives and reviewed together with its experimental prospects in Ref.~\cite{Diehl:2024bmd}, sharpens the motivation: the first Mellin moments of the chiral-odd transition GPDs are the local tensor transition form factors studied here.

A calculation of this transition in LCSR faces two difficulties. The first concerns the parametrization of $\langle \Delta | \bar q\,\sigma_{\mu\nu}\,q | N\rangle$. The matrix element requires six independent form factors, a number fixed by the dimension of the parity-reduced space of helicity amplitudes, and the six organize into transverse multipoles of rank zero to three (Sec.~\ref{sec:param} and Appendix~\ref{app:param}). The second difficulty is specific to sum rules with a spin-$3/2$ interpolating current, which couples also to spin-$1/2$ states. The standard procedure removes their contributions by discarding the Lorentz structures that carry $\gamma_\alpha$ or $p'_\alpha$ at a fixed position of an ordered Dirac string~\cite{Belyaev:1982cd, Belyaev:1993ss} and then matches, for each form factor, one of the structures that survive. The surviving structures are not unique. For the tensor transition several of them multiply the same combination of form factors on the hadronic side but differ on the QCD side, and we find that the corresponding sum rules give values that differ by up to a factor of two while being equally stable in the Borel parameter (Sec.~\ref{sec:structures}). Stability in the auxiliary parameters is therefore not sufficient to fix the result, and a criterion for the choice of the sum rule is needed.

Such a criterion is provided by the light-cone projection technique developed for the electromagnetic $N \to \Delta$ transition by Braun, Lenz, Peters and Radyushkin~\cite{Braun:2005be}. The correlation function is decomposed into components ordered by their power of the large light-cone momentum of the nucleon, only the components of leading and next-to-leading power are used, and the spin-$1/2$ contributions are eliminated by forming the combinations of these components to which spin-$1/2$ states do not contribute. Applied to the tensor transition, with spin-$1/2$ states of both parities and arbitrary masses, the method yields an overdetermined system of sum rules, so that the result can be tested internally, and it fixes which combinations of form factors the leading light-cone components can determine. We find that they determine the transverse multipoles of rank one, two and three, but neither the monopole nor the longitudinal form factor $F_1$. The single Lorentz structures that the conventional procedure would use for the same quantities have no component of leading or next-to-leading power, which accounts for their mutual inconsistency.

The analysis is restricted to the isovector tensor current. The $\Delta$ has isospin $3/2$ and the nucleon $1/2$, so that an isoscalar operator cannot connect the two states, and in the isospin limit only the combination $\bar u\,\sigma_{\mu\nu}\,u - \bar d\,\sigma_{\mu\nu}\,d$ contributes. We use two sets of nucleon distribution amplitudes (DAs)~\cite{Braun:2006hz} and take the $\Delta$ coupling and the continuum threshold from the $\Delta$ two-point sum rule in the form used in Ref.~\cite{Braun:2005be}, so that the hadronic input is fixed independently of the transition under study. The sum rules are evaluated for $1 \leq Q^2 \leq 6$~GeV$^2$.

The paper is organized as follows. Section~\ref{sec:param} summarizes the parametrization and the multipole decomposition. Section~\ref{sec:lcsr} sets up the correlation function, the light-cone projections and the removal of spin-$1/2$ contributions, determines the accessible combinations, and compares with sum rules built on single structures. Section~\ref{sec:numerical} contains the numerical analysis and Sec.~\ref{sec:summary} our conclusions. The derivation of the parametrization and of the multipole content is given in Appendix~\ref{app:param}.

\section{Parametrization and transverse multipoles}
\label{sec:param}

We write the tensor transition matrix element as
\begin{align}
&\langle \Delta(p')\,|\,\bar q\,\sigma_{\mu\nu}\,q\,|\,N(p)\rangle = \bu_\beta(p')\,\Big[\,
\frac{F_1}{\barm}\big(q_\mu g_{\beta \nu}-q_\nu g_{\beta\mu}\big) \nonumber\\
&\quad + \frac{F_2}{\barm}\big(P_\mu g_{\beta\nu}-P_\nu g_{\beta\mu}\big)
+ \frac{F_3}{\barm^2}\,q_{\beta}\big(P_\mu\gamma_\nu - P_\nu\gamma_\mu\big) \nonumber\\
&\quad + \frac{F_4}{\barm^2}\,q_{\beta}\big(q_\mu\gamma_\nu - q_\nu\gamma_\mu\big)
+ \frac{F_5}{\barm}\,i\,q_{\beta}\,\sigma_{\mu\nu} \nonumber\\
&\quad + \frac{F_6}{\barm^3}\,q_{\beta}\big(q_\mu P_\nu - q_\nu P_\mu\big)\Big]\,\gamma_5\,u_N(p),
\label{eq:TFF-param}
\end{align}
with $q = p'-p$, $Q^2 = -q^2$, $P = (p+p')/2$ and $\barm = (m_N + m_\Delta)/2$. The six form factors $F_i \equiv F_i^{N\Delta}(Q^2)$ are real and dimensionless. The trailing $\gamma_5$ is required by parity in the $\tfrac12^+ \to \tfrac32^+$ channel and occupies the same position as in the $N \to \Delta$ energy--momentum tensor parametrization~\cite{Kim:2022bia}; the factor $i$ multiplying $F_5$ is required by time reversal. The number of independent structures is fixed at six by the dimension of the parity-reduced space of helicity amplitudes, computed with the six components of $\sigma_{\mu\nu}$ included; the same counting reproduces the known form-factor content of the electromagnetic, axial and gravitational $N \to \Delta$ channels. The derivation is given in Appendix~\ref{app:param}. The choice of the six structures is not unique: one relation holds among the seven that survive the Rarita--Schwinger reduction, so that a different but equally complete basis would assign different values to the individual $F_i$.

The form factors organize into transverse multipoles. In the symmetric frame of zero skewness, $\Delta^+ = 0$, $p_\perp = -\tfrac12\Dperp$, $p'_\perp = \tfrac12\Dperp$, in which $t = -\Dperp^{\,2}$ exactly, each light-front helicity amplitude of the chiral-odd projection $\bar q\,\sigma^{+j}q$ carries a single azimuthal harmonic $e^{iM\varphi}|\Dperp|^{|M|}$, with $M = -(\lambda_\Delta - \lambda_N - s)$ and $s = \pm 1$ the transverse index of the operator in the circular basis. The rank $|M|$ is bounded by the sum of the rank of the spin transition and the rank of the operator. For $\tfrac12 \to \tfrac32$ and the tensor operator the bound is $2 + 1 = 3$, whereas the diagonal nucleon tensor sector and the electromagnetic $N \to \Delta$ transition, whose multipoles $M1$, $E2$ and $C2$ saturate their bound, stop at $|M| = 2$. Defining the multipoles from the maximal-helicity channels $\lambda_\Delta = 3/2$ and taking the leading power of $|\Dperp|$ at each rank gives
\begin{align}
\mathcal{M}_0 &= F_2, & \mathcal{M}_1 &= F_3 - F_5 + \tfrac12 F_2, \nonumber\\
\mathcal{M}_2 &= F_4 - \tfrac12 F_3 + k\,F_6, & \mathcal{M}_3 &= F_6,
\label{eq:multipoles}
\end{align}
with $k = (m_\Delta - m_N)/(m_N + m_\Delta) \simeq 0.134$; the coefficients $\pm\tfrac12$ are exact and mass independent.

The monopole plays the role of a tensor charge of the transition, the dipole is the transition analogue of the anomalous tensor magnetic moment $\kappa_T$, and the quadrupole that of $\widetilde H_T$ in the diagonal nucleon case; the octupole has no diagonal analogue, since it requires a spin transition of rank two together with a chiral-odd operator. Two further properties of Eq.~\eqref{eq:multipoles} matter below. The multipoles are labelled by a quantum number of rotations in the transverse plane and are determined by the helicity amplitudes without reference to the basis of Eq.~\eqref{eq:TFF-param}; statements about them are therefore free of the convention dependence of the individual $F_i$. And $F_1$ enters none of them: at zero skewness the structure multiplying $F_1$ acts only through the plus component of the Rarita--Schwinger field, which is suppressed in the maximal-helicity channels, so that $F_1$ is a longitudinal-polarization form factor. Since eight parity-independent amplitudes are generated by six form factors, two relations hold among the amplitudes, and the four maximal-helicity multipoles span four of the six directions in form-factor space; the remaining two are $F_1$ and $F_3 + \tfrac12 F_4 + F_5$, carried by the $\lambda_\Delta = 1/2$ channels (Appendix~\ref{app:param}). The multipoles should not be read as moments of a spatial density: for a transition the Fourier conjugate object is an interference density between two states rather than a probability.

\section{Light-cone sum rules}
\label{sec:lcsr}

\subsection{Correlation function}

We consider the correlation function
\begin{equation}
\Pi_{\alpha\mu\nu}(p,q)
= i\!\int\! d^4x\; e^{i q\cdot x}\,
\langle 0 |\,\mathcal{T}\{\, J^\Delta_\alpha(0)\,j_{\mu\nu}(x)\,\}\, | N(p)\rangle,
\label{eq:CF-def}
\end{equation}
in which the on-shell nucleon $N(p)$ is the external state, the $\Delta^+$ is created by the interpolating current
\begin{equation}
J^\Delta_\alpha = \tfrac{1}{\sqrt 3}\,\epsilon^{abc}\,
\Big[\, 2\,\big(u^{aT} C\gamma_\alpha\, d^b\big)\,u^c + \big(u^{aT} C\gamma_\alpha\, u^b\big)\,d^c\,\Big],
\label{eq:currents}
\end{equation}
with $C$ the charge-conjugation matrix and $a,b,c$ colour indices, and $j_{\mu\nu} = \bar u\,\sigma_{\mu\nu}\,u - \bar d\,\sigma_{\mu\nu}\,d$ is the isovector tensor current. The current \eqref{eq:currents} is that of Ref.~\cite{Braun:2005be} divided by $\sqrt3$.

On the hadronic side the $\Delta$ contributes
\begin{equation}
\Pi^{(\Delta)}_{\alpha\mu\nu} = \frac{\lambda_\Delta}{m_\Delta^2 - p'^2}\,\sum_{s'} u_\alpha(p',s')\,\langle \Delta(p',s')|\,j_{\mu\nu}(0)\,|N(p)\rangle,
\label{eq:Had-Delta}
\end{equation}
with $p' = p + q$, $\langle 0 | J^\Delta_\alpha(0) | \Delta(p',s')\rangle = \lambda_\Delta\,u_\alpha(p',s')$, the matrix element \eqref{eq:TFF-param}, and the Rarita--Schwinger spin sum
\begin{multline}
\sum_{s'}u_\alpha(p',s')\,\bu_{\beta}(p',s')
= -(\slashed{p}' + m_\Delta)
\Big\{ g_{\alpha\beta} - \frac{1}{3}\gamma_\alpha\gamma_{\beta} \\
- \frac{2\,p'_\alpha p'_{\beta}}{3\,m_\Delta^2}
+ \frac{p'_\alpha\gamma_{\beta} - p'_{\beta}\gamma_\alpha}{3\,m_\Delta}\Big\}.
\label{eq:RS-spin-sum}
\end{multline}
The current also couples to spin-$1/2$ states. For negative parity the coupling has the form given in Ref.~\cite{Braun:2005be}, and for positive parity we use the form that satisfies the same condition $\gamma^\alpha \langle 0|J^\Delta_\alpha|B\rangle = 0$,
\begin{align}
\langle 0 | J^\Delta_\alpha | B_-(p')\rangle &= \lambda_-\,(m_- \gamma_\alpha - 4p'_\alpha)\,u(p'), \nonumber\\
\langle 0 | J^\Delta_\alpha | B_+(p')\rangle &= \lambda_+\,(m_+ \gamma_\alpha + 4p'_\alpha)\,\gamma_5\,u(p').
\label{eq:spin12-coupling}
\end{align}
We have checked that allowing independent coefficients of $\gamma_\alpha$ and $p'_\alpha$ in Eq.~\eqref{eq:spin12-coupling} does not change the set of form-factor combinations determined below. For the transition $N \to B_\pm$ we take the most general vertex built from $\sigma_{\mu\nu}$, $P_\mu\gamma_\nu - P_\nu\gamma_\mu$, $q_\mu\gamma_\nu - q_\nu\gamma_\mu$, $P_\mu q_\nu - P_\nu q_\mu$ and their products with $\slashed{q}$, eight structures in all.

On the QCD side the correlation function is computed at leading order in $\alpha_s$ and in the chiral limit, with the free light-quark propagator and the nucleon matrix element of three quark fields at light-like separations, expanded in the nucleon DAs of twist three to six~\cite{Braun:2006hz}. Four-particle DAs are not included. The terms involving the auxiliary functions $V_1^M$, $A_1^M$ and $T_1^M$ of Ref.~\cite{Braun:2006hz}, which arise from nucleon-mass corrections, do not contribute to the combinations determined in Sec.~\ref{sec:accessible}; we have verified this numerically and omit them.

\subsection{Light-cone projections}

Following Ref.~\cite{Braun:2005be}, we introduce a light-like vector $n$ with $n\cdot q = 0$, the light-like vector $\tilde p = p - m_N^2\,n/(2\,p\cdot n)$, the transverse momentum transfer $q_\perp = q - (p\cdot q/p\cdot n)\,n$, and a unit transverse vector $e$ orthogonal to $n$, $\tilde p$ and $q_\perp$. With the projectors $\Lambda_+ = \tilde{\slashed{p}}\,\slashed{n}/(2\,p\cdot n)$ and $\Lambda_- = \slashed{n}\,\tilde{\slashed{p}}/(2\,p\cdot n)$ the nucleon spinor splits into $N_\pm = \Lambda_\pm N$. A projection contracts the free indices of $\Pi_{\alpha\mu\nu}$ with three vectors $(a,b,c)$ chosen from $\{n, e, q_\perp, \tilde p\}$, acts with $\Lambda_\pm$ from the left, and expands the result as
\begin{equation}
\Lambda_\pm\,\Pi_{\alpha\mu\nu}\,a^\alpha b^\mu c^\nu
= \sum_{i=1}^{4} \pi_i^{(\pm;\,abc)}\big(p'^2, Q^2\big)\;\Gamma_i\,N_\pm,
\label{eq:LC-proj}
\end{equation}
with $\Gamma_i \in \{\gamma_5,\ \slashed{q}_\perp\gamma_5,\ \slashed{e}\,\gamma_5,\ \slashed{e}\,\slashed{q}_\perp\gamma_5\}$.
The projections are ordered by their power of the large light-cone momentum. The two projections with $(a,b,c) = (n,n,e)$ and $(n,n,q_\perp)$ and the spinor $N_+$ are of leading power. Replacing $N_+$ by $N_-$, the transverse vector $c$ by $\tilde p$, or one of the two vectors $n$ by a transverse one lowers the power by one unit; we retain the eight projections obtained in this way, $(n,n,e)$ and $(n,n,q_\perp)$ with $N_-$, and $(n,n,\tilde p)$, $(e,n,q_\perp)$, $(q_\perp,n,e)$, $(e,n,e)$, $(q_\perp,n,q_\perp)$, $(n,e,q_\perp)$ with $N_+$. Of the forty invariant functions $\pi_i$ defined by these ten projections, twenty are non-vanishing. We collect them in a vector $\bm\pi$. The Borel transformation in $p'^2$ and the continuum subtraction are applied to each component with the prescriptions of Ref.~\cite{Braun:2006hz}, including the surface terms at the duality point $x_0$, defined by $s(x_0) = s_0$ with $s(x) = (1-x)\,m_N^2 + (1-x)\,Q^2/x$.

\subsection{Spin-1/2 contributions and accessible combinations}
\label{sec:accessible}

The $\Delta$ contribution to $\bm\pi$ is linear in the form factors, $\bm\pi^{(\Delta)} = \lambda_\Delta\,(m_\Delta^2 - p'^2)^{-1}\,H_\Delta\,\bm F$, with $\bm F = (F_1,\ldots,F_6)$ and $H_\Delta$ a $20\times6$ matrix of kinematic coefficients. Each spin-$1/2$ state contributes a vector $\bm h_B$ fixed by its mass, its parity, its vertex and Eq.~\eqref{eq:spin12-coupling}. Evaluating $\bm h_B$ for both parities, for seven masses between $1.3$ and $2.4$~GeV and for each of the eight vertex structures, we find that these vectors span an eleven-dimensional subspace, and that its dimension does not change when further masses or structures are added or when the normalization of $n$ or the external spinors are changed. Let $C$ be the $9\times20$ matrix whose rows span the orthogonal complement. The nine combinations $C\bm\pi$ receive no contribution from spin-$1/2$ states of either parity and any mass, and the sum rules read
\begin{equation}
\lambda_\Delta\,e^{-m_\Delta^2/M_B^2}\,C H_\Delta\,\bm F = C\,\mathcal{B}\big[\bm\pi^{\rm QCD}\big](Q^2; M_B^2, s_0),
\label{eq:LC-SR}
\end{equation}
where $\mathcal{B}$ denotes the Borel transformation with continuum subtraction.

The $9\times6$ matrix $C H_\Delta$ has rank four at all $Q^2$ considered. The sum rules \eqref{eq:LC-SR} therefore determine four linear combinations of the form factors and are blind to two. The multipoles $\mathcal{M}_1$, $\mathcal{M}_2$ and $\mathcal{M}_3$ lie in the determined subspace to numerical precision, whereas $\mathcal{M}_0 = F_2$ and $F_1$ have components of order one along the two undetermined directions: in those directions the $\Delta$ contribution cannot be separated from spin-$1/2$ contributions at this order of the light-cone expansion. The fourth determined combination can be written as $F_5 + b(Q^2)\,F_2$, with $b = \ensuremath{-0.18}$ at $Q^2 = 1$~GeV$^2$ and $b = \ensuremath{-0.45}$ at $Q^2 = 6$~GeV$^2$; it carries no fixed angular-momentum label and, like the individual $F_i$, depends on the basis of Eq.~\eqref{eq:TFF-param}, so we do not use it. The leading light-cone components of the correlation function thus resolve the transverse multipoles of rank one to three, and neither the monopole nor the longitudinal form factor.

Nine equations for four unknowns leave five relations that the QCD side has to satisfy if the correlation function is saturated by the $\Delta$ and spin-$1/2$ states below the continuum. These relations are not automatic: correlators of the same symmetry built from random coefficients lie almost entirely, by $96$--$98\%$ in norm, outside the subspace allowed by the $\Delta$ and spin-$1/2$ contributions. The computed correlation function satisfies them, after Borel transformation and continuum subtraction, to numerical precision at every $Q^2$, $M_B^2$ and $s_0$ considered. This tests the contractions, the projections and the Borel transformation together. The results are unchanged within numerical precision when the normalization of $n$ is varied between $0.5$ and $2.7$, when other external spinors are used, and when the set of spin-$1/2$ masses or vertex structures is enlarged. The sum rules were evaluated in two independent implementations, one built on symbolic Dirac algebra with Tracer and one numerical, which agree to numerical precision.

Adding the fourteen projections of the next power makes all six form factors formally accessible. The enlarged system, however, is not consistent: its relative residual is $38$--$43\%$, and its solution depends on the normalization of $n$, on which a physical result cannot depend. This is in line with the motivation of Ref.~\cite{Braun:2005be} for basing the sum rules on the leading projections only, and we do not use the next power.

\subsection{Comparison with sum rules from single Lorentz structures}
\label{sec:structures}

In the conventional procedure one orders the Dirac string of the correlation function as $\gamma_\alpha\gamma_\mu\gamma_\nu\slashed{q}\slashed{p}\gamma_5$, discards the structures with $\gamma_\alpha$ at the left or with $p'_\alpha$, and matches one surviving structure per form factor~\cite{Belyaev:1982cd, Belyaev:1993ss}. Two surviving structures,
\begin{align}
X_a &= q_\alpha\big(q_\nu\gamma_\mu - q_\mu\gamma_\nu\big)\gamma_5, \nonumber\\
X_b &= q_\alpha\big(q_\nu\gamma_\mu - q_\mu\gamma_\nu\big)\slashed{q}\,\gamma_5,
\label{eq:Xab}
\end{align}
both multiply, on the hadronic side, the single combination $F_3 - 2F_4$ and nothing else. By Eq.~\eqref{eq:multipoles} this combination is fixed by the multipoles, $F_3 - 2F_4 = -2\mathcal{M}_2 + 2k\,\mathcal{M}_3$, so that the light-cone sum rules give a definite value for it. Table~\ref{tab:structures} compares the three determinations with the same input and the same window (Sec.~\ref{sec:numerical}). The two single-structure sum rules are as stable in the Borel parameter as the light-cone result, but they differ from each other by a factor of about two, and from the light-cone result by a further factor of $1.5$--$2$.

\begin{table}[htb]
\centering
\small
\setlength{\tabcolsep}{5pt}
\renewcommand{\arraystretch}{1.15}
\begin{tabular}{c c c c}
\toprule
$Q^2$ [GeV$^2$] & $X_a$ & $X_b$ & light-cone \\
\midrule
1 & 1.224 (0.8\%) & 2.279 (2.1\%) & 0.789 (2.2\%) \\
2 & 0.302 (3.5\%) & 0.637 (2.6\%) & 0.175 (1.8\%) \\
3 & 0.1087 (4.9\%) & 0.2457 (4.8\%) & 0.0586 (3.6\%) \\
\bottomrule
\end{tabular}
\caption{The combination $F_3 - 2F_4$ from the single structures of Eq.~\eqref{eq:Xab} and from the light-cone sum rules, for Set-I at $M_B^2 = 2.0$~GeV$^2$ and $s_0 = 3.0$~GeV$^2$. In parentheses: the largest relative deviation for $M_B^2 = 1.5$ and $2.5$~GeV$^2$.}
\label{tab:structures}
\end{table}

The origin of the discrepancy is visible in the light-cone decomposition. Neither $X_a$ nor $X_b$ has a component in any of the ten projections of leading and next-to-leading power; both enter only at the next power, where the sum rules have been found to be inconsistent. The same holds for the structures that would be used to isolate $F_1$ and $F_5$, whereas those for $F_2$, $F_3$ and $F_6$ carry $30$--$41\%$ of their weight at next-to-leading power and the remainder beyond it. A sum rule built on a single structure of this kind is therefore controlled by the part of the light-cone expansion that is least reliable, and its Borel stability does not indicate the contrary.

\section{Numerical analysis}
\label{sec:numerical}

\begin{table}[t]
\centering
\small
\setlength{\tabcolsep}{10pt}
\renewcommand{\arraystretch}{1.15}
\begin{tabular}{l c c}
\toprule
Parameter & Set-I & Set-II \\
\midrule
$f_N$ [$10^{-3}$~GeV$^2$]         & $5.0 \pm 0.5$  & $5.0 \pm 0.5$ \\
$\lambda_1$ [$10^{-2}$~GeV$^2$]   & $-2.7 \pm 0.9$ & $-2.7 \pm 0.9$ \\
$\lambda_2$ [$10^{-2}$~GeV$^2$]   & $5.4 \pm 1.9$  & $5.4 \pm 1.9$ \\
\midrule
$A_1^u$       & $0.38 \pm 0.15$ & $0$    \\
$V_1^d$       & $0.23 \pm 0.03$ & $1/3$  \\
$f_1^d$       & $0.40 \pm 0.05$ & $3/10$ \\
$f_2^d$       & $0.22 \pm 0.05$ & $4/15$ \\
$f_1^u$       & $0.07 \pm 0.05$ & $1/10$ \\
\bottomrule
\end{tabular}
\caption{Parameters of the nucleon DAs~\cite{Braun:2006hz}. Set-I corresponds to QCD sum-rule estimates of the shape parameters, Set-II to their asymptotic values; with the latter the DAs coincide with the asymptotic expressions of Ref.~\cite{Braun:2005be}.}
\label{tab:DA-params}
\end{table}

\begin{figure*}[t]
\centering
\includegraphics[width=0.32\textwidth]{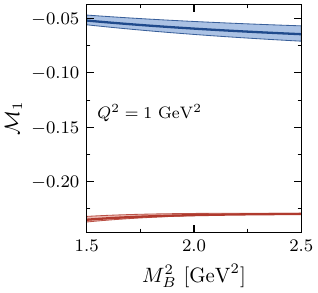}\hfill
\includegraphics[width=0.32\textwidth]{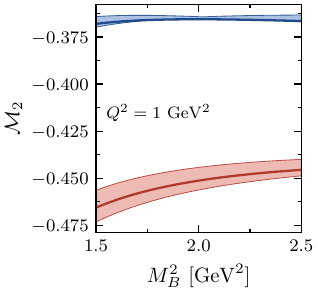}\hfill
\includegraphics[width=0.32\textwidth]{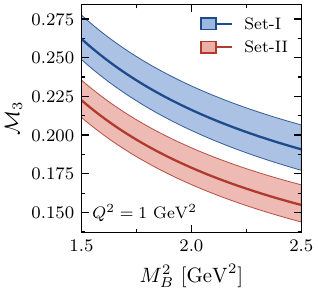}
\caption{Dependence of $\mathcal{M}_1$, $\mathcal{M}_2$ and $\mathcal{M}_3$ on the Borel parameter at $Q^2 = 1$~GeV$^2$ for Set-I (blue) and Set-II (red). The lines correspond to $s_0 = 3.0$~GeV$^2$ and the bands to $2.8 \leq s_0 \leq 3.2$~GeV$^2$, with $\lambda_\Delta$ varied together with $s_0$.}
\label{fig:borel}
\end{figure*}

\begin{figure*}[t]
\centering
\includegraphics[width=0.32\textwidth]{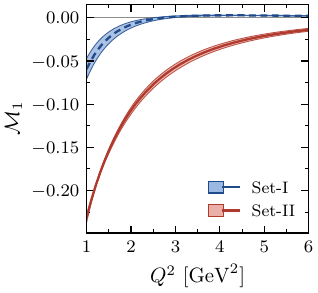}\hfill
\includegraphics[width=0.32\textwidth]{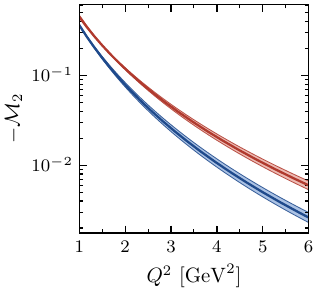}\hfill
\includegraphics[width=0.32\textwidth]{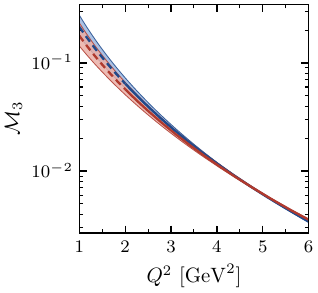}
\caption{$Q^2$ dependence of $\mathcal{M}_1$, $-\mathcal{M}_2$ and $\mathcal{M}_3$ for Set-I (blue) and Set-II (red). The lines are the central values at $M_B^2 = 2.0$~GeV$^2$ and $s_0 = 3.0$~GeV$^2$, and the bands the range on the $3\times3$ grid of $M_B^2$ and $s_0$. Dashed lines mark the results that do not meet the Borel stability criterion: $\mathcal{M}_3$ for $Q^2 < 2$~GeV$^2$, and $\mathcal{M}_1$ in Set-I, which results from a near cancellation.}
\label{fig:qsq}
\end{figure*}

\begin{table*}[t]
\centering
\small
\setlength{\tabcolsep}{8pt}
\renewcommand{\arraystretch}{1.2}
\begin{tabular}{c c c c c}
\toprule
 & $Q^2$ [GeV$^2$] & $\mathcal{M}_1$ & $\mathcal{M}_2$ & $\mathcal{M}_3$ \\
\midrule
\multirow{7}{*}{Set-I} & 1 & \ensuremath{-0.0593}\,[\ensuremath{-0.0709},\,\ensuremath{-0.0469}] & \ensuremath{-0.366}\,[\ensuremath{-0.370},\,\ensuremath{-0.363}] & 0.216\,[0.177,\,0.278] \\
 & 1.5 & \ensuremath{-0.0230}\,[\ensuremath{-0.0299},\,\ensuremath{-0.0164}] & \ensuremath{-0.158}\,[\ensuremath{-0.164},\,\ensuremath{-0.153}] & 0.112\,[0.098,\,0.134] \\
 & 2 & \ensuremath{-0.00797}\,[\ensuremath{-0.01186},\,\ensuremath{-0.00466}] & \ensuremath{-0.0789}\,[\ensuremath{-0.0838},\,\ensuremath{-0.0748}] & 0.0644\,[0.0586,\,0.0730] \\
 & 3 & 0.00134\,[0.00033,\,0.00216] & \ensuremath{-0.0259}\,[\ensuremath{-0.0283},\,\ensuremath{-0.0237}] & 0.0254\,[0.0244,\,0.0270] \\
 & 4 & 0.00282\,[0.00232,\,0.00313] & \ensuremath{-0.0106}\,[\ensuremath{-0.0118},\,\ensuremath{-0.0095}] & 0.0118\,[0.0116,\,0.0120] \\
 & 5 & 0.00264\,[0.00221,\,0.00291] & \ensuremath{-0.00501}\,[\ensuremath{-0.00564},\,\ensuremath{-0.00444}] & 0.00610\,[0.00603,\,0.00619] \\
 & 6 & 0.00216\,[0.00173,\,0.00250] & \ensuremath{-0.00262}\,[\ensuremath{-0.00297},\,\ensuremath{-0.00230}] & 0.00343\,[0.00335,\,0.00354] \\
\midrule
\multirow{7}{*}{Set-II} & 1 & \ensuremath{-0.231}\,[\ensuremath{-0.237},\,\ensuremath{-0.229}] & \ensuremath{-0.451}\,[\ensuremath{-0.473},\,\ensuremath{-0.440}] & 0.179\,[0.144,\,0.236] \\
 & 1.5 & \ensuremath{-0.154}\,[\ensuremath{-0.157},\,\ensuremath{-0.152}] & \ensuremath{-0.217}\,[\ensuremath{-0.219},\,\ensuremath{-0.215}] & 0.0966\,[0.0826,\,0.1180] \\
 & 2 & \ensuremath{-0.107}\,[\ensuremath{-0.111},\,\ensuremath{-0.104}] & \ensuremath{-0.119}\,[\ensuremath{-0.123},\,\ensuremath{-0.117}] & 0.0572\,[0.0510,\,0.0664] \\
 & 3 & \ensuremath{-0.0572}\,[\ensuremath{-0.0613},\,\ensuremath{-0.0535}] & \ensuremath{-0.0451}\,[\ensuremath{-0.0479},\,\ensuremath{-0.0426}] & 0.0238\,[0.0224,\,0.0259] \\
 & 4 & \ensuremath{-0.0335}\,[\ensuremath{-0.0367},\,\ensuremath{-0.0306}] & \ensuremath{-0.0206}\,[\ensuremath{-0.0224},\,\ensuremath{-0.0190}] & 0.0115\,[0.0112,\,0.0120] \\
 & 5 & \ensuremath{-0.0210}\,[\ensuremath{-0.0234},\,\ensuremath{-0.0189}] & \ensuremath{-0.0106}\,[\ensuremath{-0.0117},\,\ensuremath{-0.0096}] & 0.00615\,[0.00611,\,0.00625] \\
 & 6 & \ensuremath{-0.0139}\,[\ensuremath{-0.0156},\,\ensuremath{-0.0123}] & \ensuremath{-0.00598}\,[\ensuremath{-0.00667},\,\ensuremath{-0.00534}] & 0.00355\,[0.00352,\,0.00361] \\
\bottomrule
\end{tabular}
\caption{Transverse multipoles of the isovector $N \to \Delta$ tensor transition: central value at $M_B^2 = 2.0$~GeV$^2$, $s_0 = 3.0$~GeV$^2$, and in brackets the range on the $3\times3$ grid $M_B^2 \in \{1.5, 2.0, 2.5\}$~GeV$^2$, $s_0 \in \{2.8, 3.0, 3.2\}$~GeV$^2$.}
\label{tab:results}
\end{table*}

\subsection{Input}

We use $m_N = 0.94$~GeV and $m_\Delta = 1.232$~GeV and the two sets of nucleon DA parameters of Ref.~\cite{Braun:2006hz} listed in Table~\ref{tab:DA-params}: Set-I, based on QCD sum-rule estimates of the shape parameters, and Set-II, the asymptotic values. The coupling $\lambda_\Delta$ and the continuum threshold $s_0$ are taken from the Belyaev--Ioffe sum rule for the correlation function of two $\Delta$ currents~\cite{Belyaev:1982cd}, in the form of Eq.~(C.1) of Ref.~\cite{Braun:2005be},
\begin{multline}
\frac13\,\tilde\lambda_\Delta^2\,e^{-m_\Delta^2/M^2}
= \frac{M^6}{5}\Big[1 - e^{-s_0/M^2}\Big(1 + \frac{s_0}{M^2} + \frac{s_0^2}{2M^4}\Big)\Big] \\
- \frac{5b}{72}\,M^2\big(1 - e^{-s_0/M^2}\big) + \frac{4a^2}{3} - \frac{7m_0^2a^2}{9M^2},
\label{eq:twopoint}
\end{multline}
with $a = -(2\pi)^2\langle\bar qq\rangle = 0.55$~GeV$^3$, $b = (2\pi)^2\langle\alpha_s G^2/\pi\rangle = 0.47$~GeV$^4$ and $m_0^2 = 0.8$~GeV$^2$~\cite{Braun:2005be}. Here $\tilde\lambda_\Delta$ refers to the current of Ref.~\cite{Braun:2005be} and the normalization $\langle 0|\eta_\mu|\Delta\rangle = \tilde\lambda_\Delta/(2\pi)^2\,u_\mu$, so that for the current \eqref{eq:currents} $\lambda_\Delta = \tilde\lambda_\Delta/[\sqrt3\,(2\pi)^2]$. The sum rule \eqref{eq:twopoint} varies by less than $4\%$ over $1.5 \leq M^2 \leq 2.5$~GeV$^2$ for $2.8 \leq s_0 \leq 3.2$~GeV$^2$, the interval $s_0 = 3.0 \pm 0.2$~GeV$^2$ found in Ref.~\cite{Braun:2005be}. Averaging over this range of $M^2$ gives $\tilde\lambda_\Delta = 1.84$, $1.91$ and $1.97$~GeV$^3$ for $s_0 = 2.8$, $3.0$ and $3.2$~GeV$^2$, consistent with $\tilde\lambda_\Delta = 2.0 \pm 0.2$~GeV$^3$ of Ref.~\cite{Braun:2005be}, that is $\lambda_\Delta = 0.0270$, $0.0279$ and $0.0288$~GeV$^3$. In the light-cone sum rules $\lambda_\Delta$ is varied together with $s_0$. The two-point sum rule determines only $\lambda_\Delta^2$; we take $\lambda_\Delta > 0$. With the opposite choice all form factors, and hence all multipoles, change sign together, so that only their relative signs are physical.

The Borel parameter of the light-cone sum rules is taken in the interval $1.5 \leq M_B^2 \leq 2.5$~GeV$^2$ with the central value $2.0$~GeV$^2$, as in Ref.~\cite{Braun:2005be}. The lower end is set by the onset of stability of the two-point sum rule and the upper end by the convergence of the light-cone expansion, whose effective parameter is of order $1/(\langle x\rangle M_B^2)$ rather than $1/M_B^2$. As criterion we require that a result vary by less than $10$--$15\%$ within this interval, the stability that Ref.~\cite{Braun:2005be} found for the electromagnetic transition. The window is fixed at $Q^2 = 1$~GeV$^2$ and used unchanged over $1 \leq Q^2 \leq 6$~GeV$^2$. We quote central values at $M_B^2 = 2.0$~GeV$^2$ and $s_0 = 3.0$~GeV$^2$, and as window uncertainty the range spanned on the $3\times3$ grid $M_B^2 \in \{1.5, 2.0, 2.5\}$~GeV$^2$, $s_0 \in \{2.8, 3.0, 3.2\}$~GeV$^2$. Taking the range rather than a sum in quadrature avoids an assumption on the correlation of the two sources. The difference between the two DA sets is shown separately.

\subsection{Results}

Figure~\ref{fig:borel} shows the three multipoles as functions of $M_B^2$ at $Q^2 = 1$~GeV$^2$, Fig.~\ref{fig:qsq} their dependence on $Q^2$, and Table~\ref{tab:results} the values at selected momentum transfers.

Measured against the criterion stated above, the quadrupole is the most stable of the three. Its Borel variation stays below 6.4\% in Set-I and 6.0\% in Set-II over the whole interval $1 \leq Q^2 \leq 6$~GeV$^2$, and its variation over the $s_0$ band below 7.8\% and 6.5\%. At $Q^2 = 1$~GeV$^2$ one finds $\mathcal{M}_2 = \ensuremath{-0.366}$ for Set-I and $\ensuremath{-0.451}$ for Set-II. The model of the DAs is the dominant uncertainty, and its effect grows with $Q^2$: the ratio of the two sets is 1.23 at $Q^2 = 1$~GeV$^2$ and 2.28 at $Q^2 = 6$~GeV$^2$, Set-II falling more slowly. A growing sensitivity to the shape of the DAs is consistent with the increase of the effective momentum fraction with $Q^2$ noted for the electromagnetic sum rules in Ref.~\cite{Braun:2005be}.

The octupole behaves in the opposite way. It is the multipole least sensitive to the DAs: the two sets differ by 17\% at $Q^2 = 1$~GeV$^2$ and agree within $5\%$ for $Q^2 \geq 3.5$~GeV$^2$. At low $Q^2$, however, it does not meet the stability criterion: its Borel variation is 21\% (Set-I) and 24\% (Set-II) at $Q^2 = 1$~GeV$^2$, falls below $15\%$ at $Q^2 \simeq 1.5$--$1.75$~GeV$^2$ and below $10\%$ at $Q^2 \simeq 2$--$2.25$~GeV$^2$, and stays below 3\% for $Q^2 \geq 4$~GeV$^2$. We therefore regard $\mathcal{M}_3$ as determined for $Q^2 \gtrsim 2$~GeV$^2$. The octupole falls more slowly with $Q^2$ than the quadrupole and exceeds $|\mathcal{M}_2|$ in Set-I for $Q^2 \gtrsim 3.2$~GeV$^2$.

The dipole is determined with a small window uncertainty in each DA set separately but depends strongly on the set. In Set-II its Borel and $s_0$ variations stay below 6\% and 7\% over the whole interval, with $\mathcal{M}_1 = \ensuremath{-0.231}$ at $Q^2 = 1$~GeV$^2$. In Set-I it results from a near cancellation between the contributions of $F_3$, $F_5$ and $F_2$ in Eq.~\eqref{eq:multipoles}: at $Q^2 = 1$~GeV$^2$ it is $\ensuremath{-0.059}$, smaller in magnitude than in Set-II by a factor 3.9, its Borel variation exceeds $12\%$ for $Q^2 \leq 2.5$~GeV$^2$, and it changes sign near $Q^2 \simeq 2.7$~GeV$^2$. The present analysis therefore fixes the dipole as negative for $Q^2 \lesssim 2.5$~GeV$^2$, with a magnitude that depends on the DAs: at $Q^2 = 1$~GeV$^2$ the ratio $\mathcal{M}_1/\mathcal{M}_2$ is 0.16 for Set-I and 0.51 for Set-II.

All three multipoles fall steeply with $Q^2$: between $Q^2 = 1$ and $2$~GeV$^2$ the quadrupole decreases by a factor 4.6 (Set-I) and 3.8 (Set-II), and the octupole by 3.4 and 3.1. The steep fall is a property of the sum rules for this correlation function rather than of the light-cone projections: it is present also in the single-structure sum rules of Table~\ref{tab:structures}, whose values decrease by factors of 3.6 and 4.1 between $Q^2 = 1$ and $2$~GeV$^2$.

\subsection{The static limit}

The light-cone sum rules apply for $Q^2 \gtrsim 1$~GeV$^2$, and the values at $Q^2 = 0$ require an extrapolation. We have tested whether the results constrain it by fitting the central values on $1 \leq Q^2 \leq 6$~GeV$^2$ with three forms that are commonly used: a pole with free power, $f_0\,[1 + Q^2/(\alpha\mathcal{M}^2)]^{-\alpha}$; the $z$ expansion to third order with $t_c = 4m_\pi^2$; and a $\rho$ pole multiplying a second-order $z$ expansion. All three describe the results with a relative root-mean-square deviation below 2.3\%. Their values at $Q^2 = 0$ are between \ensuremath{-3.6} and \ensuremath{-14.0} for $\mathcal{M}_2$ and between 1.1 and 5.5 for $\mathcal{M}_3$ in Set-I; in Set-II they are between \ensuremath{-3.2} and \ensuremath{-14.1} and between 0.82 and 3.9, and for $\mathcal{M}_1$ in Set-II they range from \ensuremath{-0.05} to \ensuremath{-0.6}, more than an order of magnitude. The extrapolation is thus fixed by the choice of the form and not by the sum rules, which is to be expected since the first $t$-channel singularity of the isovector tensor current lies at $4m_\pi^2$, far below the region in which the sum rules apply. We therefore do not quote values at $Q^2 = 0$. For the comparison with GPDs this is not a limitation: the first Mellin moments of the transition GPDs are functions of $t$, and Table~\ref{tab:results} gives them directly at $-t = Q^2$.

\section{Summary and outlook}
\label{sec:summary}

The matrix element of the tensor current between the nucleon and the $\Delta(1232)$ requires six form factors, which organize into transverse multipoles of rank zero to three; the octupole has no counterpart in the diagonal tensor sector or in the electromagnetic $N \to \Delta$ transition. We have computed the isovector transition within QCD light-cone sum rules, constructing the sum rules from the light-cone projections of leading and next-to-leading power of the correlation function and removing the contributions of spin-$1/2$ states of both parities explicitly. At this order the sum rules determine four combinations of the form factors, among which the dipole, the quadrupole and the octupole, and they obey five consistency relations to numerical precision. The monopole and the longitudinal form factor $F_1$ are not accessible: in their directions the $\Delta$ contribution cannot be separated from spin-$1/2$ contributions with the leading light-cone components, and including the next power gives an inconsistent system. Sum rules built on single Lorentz structures that measure the same combination of form factors disagree with each other by a factor of two while being equally stable in the Borel parameter; the structures involved have no component of leading or next-to-leading power. For spin-$3/2$ transitions in which several surviving structures carry the same combination of form factors, Borel stability alone does not therefore select a result.

With the $\Delta$ coupling and continuum threshold taken from the $\Delta$ two-point sum rule and two sets of nucleon DAs, the quadrupole is determined over $1 \leq Q^2 \leq 6$~GeV$^2$ with Borel and threshold variations below ten percent, and the octupole for $Q^2 \gtrsim 2$~GeV$^2$, where it is nearly independent of the DAs. The dipole is negative at $Q^2 \lesssim 2.5$~GeV$^2$, and its magnitude depends strongly on the DAs. The model of the DAs is the dominant source of uncertainty for the dipole and the quadrupole. The results do not determine an extrapolation to $Q^2 = 0$, and we do not quote static values.

The multipoles are the quantities to compare with the first Mellin moments of the chiral-odd $N \to \Delta$ transition GPDs. Since they are labelled by transverse angular momentum and do not depend on the choice of Lorentz basis, a GPD parametrization organized by that quantum number can be matched to them rank by rank; the two relations among the light-front amplitudes noted in Sec.~\ref{sec:param} are constraints on such a parametrization at the level of first moments. Lattice QCD computes the same local matrix element and can form the same combinations, including the monopole and $F_1$, which the present sum rules do not reach. Within the sum-rule approach, the dependence on the DAs is the point at which an improvement is most needed, and the same light-cone construction can be applied to the tensor and gravitational transitions of the octet baryons to the decuplet.

\appendix

\section{Lorentz parametrization and multipole content}
\label{app:param}

\subsection{Setting}
\label{app:setting}

We write the transition matrix element as
\begin{equation}
\langle \Delta(p')\,|\,\bar q\,\sigma_{\mu\nu}\,q\,|\,N(p)\rangle
= \bu_\alpha(p')\,\Gamma^{\alpha}{}_{\mu\nu}(P,q)\,u(p),
\label{eq:app-start}
\end{equation}
with $P=(p+p')/2$, $q=p'-p$, $Q^2=-q^2$ and $\barm=(m_N+m_\Delta)/2$. The Dirac kernel $\Gamma^{\alpha}{}_{\mu\nu}$ is built from $\{P,q,\gamma_\mu,g_{\mu\nu}\}$ and is antisymmetric in $(\mu\nu)$. On shell,
\begin{align}
&(\slashed{p}-m_N)\,u=0,\qquad \bu_\alpha\,(\slashed{p}'-m_\Delta)=0, \nonumber\\
&\gamma^\alpha\,\bu_\alpha=0,\qquad p'^\alpha\,\bu_\alpha=0,
\label{eq:app-onshell}
\end{align}
and, using $P^\alpha=p'^\alpha-q^\alpha/2$,
\begin{equation}
\bu_\alpha\,P^\alpha=-\tfrac12\,\bu_\alpha\,q^\alpha .
\label{eq:app-Pq}
\end{equation}

\subsection{Independent structures}
\label{app:count}

Lorentz covariance and antisymmetry in $(\mu\nu)$ admit eleven kernels:
\begin{equation}
\begin{aligned}
T^{(1)} &= g^{\alpha}{}_{\mu}q_\nu-g^{\alpha}{}_{\nu}q_\mu, &
T^{(2)} &= g^{\alpha}{}_{\mu}P_\nu-g^{\alpha}{}_{\nu}P_\mu,\\
T^{(3)} &= g^{\alpha}{}_{\mu}\gamma_\nu-g^{\alpha}{}_{\nu}\gamma_\mu, &
T^{(4)} &= q^{\alpha}\sigma_{\mu\nu},\\
T^{(5)} &= P^{\alpha}\sigma_{\mu\nu}, &
T^{(6)} &= q^{\alpha}(\gamma_\mu q_\nu-\gamma_\nu q_\mu),\\
T^{(7)} &= q^{\alpha}(\gamma_\mu P_\nu-\gamma_\nu P_\mu), &
T^{(8)} &= P^{\alpha}(\gamma_\mu q_\nu-\gamma_\nu q_\mu),\\
T^{(9)} &= P^{\alpha}(\gamma_\mu P_\nu-\gamma_\nu P_\mu), &
T^{(10)} &= q^{\alpha}(q_\mu P_\nu-q_\nu P_\mu),\\
T^{(11)} &= P^{\alpha}(q_\mu P_\nu-q_\nu P_\mu), &&
\end{aligned}
\label{eq:app-eleven}
\end{equation}
where the indices $(\alpha;\mu\nu)$ of $T^{(i)\alpha}{}_{\mu\nu}$ are suppressed on the left-hand sides.
Two further families are admissible in principle, the Levi-Civita duals such as $q^\alpha\varepsilon_{\mu\nu\rho\sigma}q^\rho P^\sigma$ and structures carrying an additional $\slashed{q}$ in the Dirac string. With the appropriate assignment of $\gamma_5$ both lie in the same parity sector as \eqref{eq:app-eleven}, by $\gamma_5\sigma_{\mu\nu}=\tfrac{i}{2}\varepsilon_{\mu\nu\rho\sigma}\sigma^{\rho\sigma}$. We have constructed them and verified that each reduces on shell to a combination of the eleven above, so that \eqref{eq:app-eleven} spans the full space.

Equation~\eqref{eq:app-Pq} converts each $P^{\alpha}$-headed kernel into its $q^{\alpha}$ partner,
\begin{align}
\bu_\alpha T^{(5)}&=-\tfrac12\,\bu_\alpha T^{(4)}, & \bu_\alpha T^{(8)}&=-\tfrac12\,\bu_\alpha T^{(6)}, \nonumber\\
\bu_\alpha T^{(9)}&=-\tfrac12\,\bu_\alpha T^{(7)}, & \bu_\alpha T^{(11)}&=-\tfrac12\,\bu_\alpha T^{(10)},
\label{eq:app-rsred}
\end{align}
leaving $T^{(1)},T^{(2)},T^{(3)},T^{(4)},T^{(6)},T^{(7)},T^{(10)}$.

The Gordon identity at unequal masses and the Clifford identity relate these seven to one another, but both follow from the Dirac equations together with the Clifford algebra and each undoes the rewriting produced by the other; they therefore do not fix the number of independent structures. That number is the dimension of the parity-reduced helicity amplitude space.

With $\mathbf q$ along $z$, the tensor operator decomposes under rotations about that axis into $\sigma_{0i}$ and $\tfrac12\epsilon_{ijk}\sigma_{jk}$, each carrying $J_z=0,\pm1$ and together exhausting the six components of $\sigma_{\mu\nu}$. Conservation of $J_z$ requires $\lambda_\Delta-\lambda_N=J_z^{\rm(op)}$, which with $\lambda_N=\pm\tfrac12$ and $\lambda_\Delta=\pm\tfrac12,\pm\tfrac32$ gives four amplitudes for each of the three values of $J_z^{\rm(op)}$, hence twelve, reduced by parity to
\begin{equation}
N_{\rm ff}=6 .
\label{eq:app-count}
\end{equation}
The six components of the operator enter this count. A counting that assigns one amplitude to each hadron helicity configuration gives $2\times4=8$, halved by parity to four, and underestimates the dimension of the amplitude space for an operator of six components.

The same rule reproduces the established content of the related channels. For $\gamma^*N\to\Delta$ it gives four, reduced to three by current conservation, matching $M1$, $E2$ and $C2$; for the axial transition four, matching $C_3^A\ldots C_6^A$, with no reduction since the axial current is not conserved; for the $N\to\Delta$ energy--momentum tensor nine, matching the five conserved and four non-conserved form factors of Ref.~\cite{Kim:2022bia}; and for the diagonal $\Delta\to\Delta$ tensor transition ten~\cite{Asmaee:2025elo, Asmaee:2026wrk}. The tensor current is not conserved, so \eqref{eq:app-count} stands.

Exactly one relation therefore holds among the seven survivors of \eqref{eq:app-rsred}. The sets
\begin{align*}
&\{T^{(1)},T^{(2)},T^{(4)},T^{(6)},T^{(7)},T^{(10)}\}, \\
&\{T^{(1)},T^{(2)},T^{(3)},T^{(6)},T^{(7)},T^{(10)}\}
\end{align*}
are complete, whereas $\{T^{(1)},T^{(2)},T^{(3)},T^{(4)},T^{(6)},T^{(7)}\}$ has rank five: $T^{(10)}$ cannot be traded against any combination of the others and must be retained. We adopt the first set.

\subsection{Parity and the parametrization}
\label{app:parity}

For $\tfrac12^+\to\tfrac32^+$ the kernel obeys
\begin{equation}
\Gamma_{\alpha\mu\nu}(P,q)=-\gamma^0\,\Gamma_{\bar\alpha\bar\mu\bar\nu}(\bar P,\bar q)\,\gamma^0,
\qquad \bar v^\mu\equiv(v^0,-\mathbf v),
\label{eq:app-parity}
\end{equation}
the overall sign reflecting the intrinsic parity of the $\Delta$ relative to the nucleon. A kernel assembled from $\{P,q,\gamma_\mu,g\}$ alone fails this condition. Taking $T^{(1)}$ with both free indices spatial, the combination $q_i g^{\alpha}{}_j-q_j g^{\alpha}{}_i$ changes sign under $\mathbf q\to-\mathbf q$ while the conjugation acts trivially on a $c$-number in Dirac space, so the two sides of \eqref{eq:app-parity} differ by an overall sign; the same holds for each of the six kernels. A single $\gamma_5$ appended on the right supplies the missing sign through $\gamma^0\gamma_5\gamma^0=-\gamma_5$, and no other placement is compatible with \eqref{eq:app-onshell}. Unlike the diagonal $\tfrac12^+\to\tfrac12^+$ case the factor is not conventional here: it originates in the spin-$1$ polarization of the Rarita--Schwinger spinor, uncompensated when only one such spinor appears. The same factor occupies the same position in the $N\to\Delta$ energy--momentum tensor parametrization~\cite{Kim:2022bia}.

The kernels have dimensions $[\text{mass}]^{1}$, $[\text{mass}]^{2}$ and $[\text{mass}]^{3}$; dividing by the corresponding power of $\barm$ renders the form factors dimensionless. The symmetric scale $\barm$ treats the initial and final states alike and matches the convention of the energy--momentum tensor analyses. The factor $i$ multiplying $q^\alpha\sigma_{\mu\nu}$ is likewise fixed rather than conventional: that structure transforms under time reversal with the opposite sign to the other five, so without the $i$ its coefficient would be purely imaginary, in conflict with the reality of the form factors. The same convention appears as $i\sigma_{\mu\nu}H_T$ in the diagonal nucleon tensor parametrization. Collecting the basis, the trailing $\gamma_5$ and the dimensional normalization gives Eq.~\eqref{eq:TFF-param}.

\subsection{Transverse multipoles}
\label{app:multipole}

Work in the symmetric frame of zero skewness,
\begin{align}
&\Delta^+=0,\qquad t=-\Dperp^{\,2}, \nonumber\\
&p_\perp=-\tfrac12\Dperp,\qquad p'_\perp=+\tfrac12\Dperp,
\label{eq:app-frame}
\end{align}
where the last relation is exact. In a three-dimensional Breit frame the mass difference enters the relation between $t$ and the momentum transfer and the point of vanishing three-momentum lies at $t=(m_\Delta-m_N)^2$; in \eqref{eq:app-frame} the mass difference drops out. Let $s=\pm1$ label the transverse index of $\bar q\,\sigma^{+j}q$ in the circular basis.

Under a rotation by $\varphi$ about the $z$ axis the states pick up $e^{-i\lambda_N\varphi}$ and $e^{+i\lambda_\Delta\varphi}$, the operator index picks up $e^{-is\varphi}$, and $\Dperp$ rotates in the plane. Invariance then fixes the azimuthal dependence of each light-front helicity amplitude to a single harmonic,
\begin{equation}
A^{(s)}_{\lambda_\Delta\lambda_N}(\Dperp)
= e^{iM\varphi}\,|\Dperp|^{|M|}\,f_{M}\big(\Dperp^{\,2}\big),
\label{eq:app-harmonic}
\end{equation}
with $M=-(\lambda_\Delta-\lambda_N-s)$.
The power $|\Dperp|^{|M|}$ follows because $\Dperp$ is the only source of azimuthal dependence, so that $|M|$ units of azimuthal angular momentum require $|M|$ powers of the transverse momentum; $f_M$ is azimuthally trivial and supplies the subleading powers $|\Dperp|^{|M|+2},|\Dperp|^{|M|+4},\ldots$ within the same rank. The integer $M$ is the multipole order.

Equation~\eqref{eq:app-harmonic} bounds the rank. The helicity flip is limited by the spin content of the transition and $s$ by the transverse index of the operator, so
\begin{equation}
|M|_{\max}=r_{\rm spin}+r_{\rm op},
\label{eq:app-ceiling}
\end{equation}
where $r_{\rm spin}$ is the rank of the spin transition and $r_{\rm op}$ that of the operator.
For $\tfrac12\otimes\tfrac12=0\oplus1$ the spin rank is one and for $\tfrac12\otimes\tfrac32=1\oplus2$ it is two; the vector and axial-vector currents carry operator rank zero in this sense while the tensor operator, whose transverse index supplies one unit, carries rank one. Hence $|M|_{\max}=2$ for the diagonal nucleon tensor sector, which is why a quadrupole exists there and is absent in the chiral-even case, and $|M|_{\max}=2$ for the electromagnetic $N\to\Delta$ transition, whose $M1$, $E2$ and $C2$ saturate that bound. For the chiral-odd $N\to\Delta$ transition $|M|_{\max}=3$.

At each rank the channel with maximal $\Delta$ helicity, $\lambda_\Delta=3/2$, carries a single power of $|\Dperp|$, whereas the $\lambda_\Delta=1/2$ channels mix leading and subleading powers. Defining the multipoles from the maximal-helicity channels, one per rank, and taking the leading power of \eqref{eq:app-harmonic} gives Eq.~\eqref{eq:multipoles}. The coefficients $\tfrac12$ and $-\tfrac12$ are exact and mass independent; the only mass dependence is the ratio $k=(m_\Delta-m_N)/(m_N+m_\Delta)\simeq0.134$ multiplying $F_6$ in $\mathcal{M}_2$.

Two features of Eq.~\eqref{eq:multipoles} follow from \eqref{eq:app-frame}. Both $F_1$ and $F_2$ multiply kernels carrying the Rarita--Schwinger index on the metric, so that $\bu_\beta$ contracts directly onto a free tensor index. Projecting onto the chiral-odd component, $(\mu,\nu)\to(+,j)$, and using $q^+=\Delta^+=0$ together with $P_\perp=0$,
\begin{align}
F_1&:\quad q^+\bu^{\,j}-q^{\,j}\bu^{+}=-\Delta^{j}\,\bu^{+}, \nonumber\\
F_2&:\quad P^+\bu^{\,j}-P^{\,j}\bu^{+}=P^{+}\,\bu^{\,j}.
\label{eq:app-F1F2}
\end{align}
The form factor $F_2$ therefore enters through the transverse component $\bu^{\,j}$ multiplied by the large light-cone momentum $P^+$, which is why it saturates the monopole. The form factor $F_1$ can enter only through $\bu^{+}$, and for $\lambda_\Delta=3/2$ the Rarita--Schwinger spinor is built from the transverse polarization $\varepsilon(m=+1)$, whose plus component is $O(1)$ rather than $O(p^+)$. The $F_1$ contribution to the maximal-helicity channels is thus suppressed by one power of $p^+$ and vanishes in the light-front limit: reaching $F_1$ requires the $m=0$ polarization vector, which enters only for $\lambda_\Delta=\pm1/2$.

Finally, the decomposition is not a change of basis. Counting independent directions rank by rank gives two at $|M|=0$, three at $|M|=1$, two at $|M|=2$ and one at $|M|=3$, so that eight parity-independent light-front amplitudes are generated by six form factors and two relations hold among them. These are constraints that a classification of chiral-odd $N\to\Delta$ transition GPDs must satisfy at the level of first Mellin moments; they are the transition analogue of the vanishing first moment of $\widetilde E_T$ in the diagonal case. Consistently, the four maximal-helicity multipoles span four of the six directions, the two they do not reach being $F_1$ and the combination $F_3+\tfrac12F_4+F_5$, both carried by the $\lambda_\Delta=1/2$ channels.

We note in closing that for a transition $Q^2=0$ is a physical point but not the point of vanishing three-momentum transfer, which lies at $t=(m_\Delta-m_N)^2$. This is the kinematic subtlety underlying Siegert's theorem in the electromagnetic $N\to\Delta$ transition.

\bibliographystyle{apsrev4-1}
\bibliography{NtoDeltaTFFs}

\end{document}